\documentclass[twocolumn,aps,showpacs]{revtex4}

\newcommand{\be}{\begin{equation}}
\newcommand{\ee}{\end{equation}}
\newcommand{\bea}{\begin{eqnarray}}
\newcommand{\eea}{\end{eqnarray}}

\begin{document}

\author{Li-Xiang Cen and Paolo Zanardi }
\affiliation{Institute for Scientific Interchange Foundation, 
Viale Settimio Severo 65, I-10133 Torino, Italy}

\title{Refocusing schemes for holonomic quantum computation
in  presence of dissipation}

\begin{abstract}
The effects of dissipation on a holonomic quantum computation 
scheme are analyzed within the quantum-jump approach. We extend 
to the  non-Abelian case the refocusing strategies formerly introduced for
(Abelian) geometric computation. We show how double loop symmetrization schemes 
allow one to get  rid of the unwanted influence of dissipation in the no-jump trajectory.
\end{abstract}

\maketitle
Due to the fragility of quantum coherence in the presence of noise, 
decoherence is the main obstacle to the practical realization of 
quantum computation. Exploring potential ways to implement robust 
quantum computation therefore is a crucial and attractive challenge.
To the aim of stabilizing quantum information a variety of
decoherence-reduction techniques have been developed, such as quantum error
correcting codes \cite{Shor:95,Knill:97b,Gottesman:97}, decoherence-free
subspaces \cite{Zanardi:97a,LidarWhaley:03} and generalized decoupling 
techniques \cite{bang}.

The geometric/holonomic approach to quantum computation 
is believed to provide an intriguing geometrical way to protect 
quantum information processing \cite{zanardi,pachos,jones,falci,duan,recati,cen}. 
Geometric quantum effects, either the simple Abelian phase 
factor \cite{berry} or its general non-Abelian counterpart 
\cite{wilzek}, arise when a quantum system is 
adiabatically driven to undergo an appropriate cyclic evolution.
A purely geometric transformation can be obtained via a refocusing 
scheme \cite{jones} which cancels out the dynamical phase or by 
properly designing the system levels \cite{zanardi,duan} 
to avoid the dynamical phase directly.

The robustness of geometric quantum computation has been 
intensively addressed recently \cite{nazir,blais,chiara,carollo,carollo2,guridi}. 
For a classical noise, the inherent stability of Berry phase
has been clearly demonstrated by studying a spin one-half system 
subject to a randomly fluctuating magnetic field \cite{chiara}.

In this paper we will consider geometric manipulations, Abelian and 
non Abelian, in presence of quantum dissipation. We will first 
analyze, the quantum-jump formalism, the non-unitary effects induced by
dissipation on the geometrical gates in order to devise strategies to 
get rid of them. We will show that the geometric gate operation based on 
a double-loop refocusing scheme does possess a certain resilience against
dissipation induced distortion. The main contribution of this paper will 
consist of  a novel  scheme which further extends  the refocusing technique to 
enact non-Abelian holonomic quantum gates by means of a sort of simultaneous 
double-loop control process.

{\em Influence of dissipation on geometric quantum computation.}
Let us consider the dissipative dynamics of a time-dependent periodic system 
governed by the following master equation 
\begin{equation}
\dot{\rho}=-i[H(t),\rho ]-\frac 12\sum_{k=1}^n(\Gamma _k^
{\dagger }\Gamma_k\rho +\rho \Gamma _k^{\dagger }\Gamma _k-
2\Gamma _k\rho \Gamma _k^{\dagger}).
\label{master}
\end{equation}
In the framework of quantum jump approach \cite{jump}, the operator 
$\Gamma _k$ is associated with jump in the trajectory; the 
time evolution of the system is calculated by averaging over 
all the possible trajectories. Stochastic jumps and diffusion of 
trajectories will destroy quantum coherence. This undesired effect becomes 
particularly severe for slow adiabatic evolutions since decoherence 
has a long time to bite in the process. 

As one could intuitively anticipate, the geometric contributions in the evolution 
will show dissipation-related aspects distinct from the dynamical evolution
ones. A promising approach to address this issue is to foucs on the evolution of 
single trajectories. In this way one can resort to the theory of geometric phases 
developed for pure states. Individual trajectories can be realized conditionally to 
the occurrence of a suitable chain of detections. Notice that this approach 
has been recently employed to calculate the geometric phase of open systems 
\cite{carollo,carollo2}.

A single trajectory in the quantum jump model is specified, for a continuously 
measured system, as a chain of states obtained by the action of a specific sequence 
of detected jumps. The state evolution conditioned to the detection of no jumps, i.e., 
the no-jump trajectory, can be described by the Schr\"{o}dinger equation 
\begin{equation}
i\frac \partial {\partial t}|\psi (t)\rangle =
\tilde{H}(t)|\psi (t)\rangle ,
\label{schro}
\end{equation}
with the non-Hermitian effective Hamiltonian given by
\begin{equation}
\tilde{H}(t)=H(t)-\frac i2\sum_{k=1}^n\Gamma _k^{\dagger }
\Gamma _k.
\label{complexh}
\end{equation}
Here, the last term contained in Eq. (\ref{complexh}) accounts for
the dissipative interaction. The state evolution indicated above 
therefore differs from the conventional unitary dissipation-free process.
This deviation then results in an inevitable distortion for the 
desired gate operations.

Let us first examine the Berry phase shift achieved through the 
double-loop refocusing scheme e.g., in an NMR system \cite{jones}. 
A spin half nucleus is assumed to be subjected to a cyclic field evolving 
adiabatically with cone angle $\theta $ and its
Hamiltonian is written as $H_0^C(t)=\frac \Omega 2(\cos \theta \sigma
_z+\sin \theta (\cos \gamma t\sigma _x+\sin \gamma t\sigma _y))$. 
The dynamical evolution will generate a dynamical phase $\phi _{\pm}^d=
\pm \frac 12\Omega T$ as well as a geometric Berry phase $\phi _{\pm }^g=
\pm \pi (1-\cos \theta )$. To remove the dynamical phase, 
the cyclic process is run twice; the second loop is surrounded by a pair 
of hard $\pi $ pulses and it is driven in the opposite orientation of the first one.
A pure geometric phase shift is then obtained. 
We consider now the system subjected to dissipation with
$\Gamma =\sqrt{\kappa }\sigma _{-}$. 
The conditional no-jump evolution of 
the system is governed by the Schr\"{o}dinger 
equation with an effective Hamiltonian (\ref{complexh}). 
Instead of a real dynamical phase originated from the unitary system, 
dissipation will give rise to a complex ``dynamical phase'' given by 
\begin{equation}
\tilde{\phi}_{\pm }^d=\pm \frac 12\bar{\Omega}T-\frac i4\kappa T, ~~
\bar{\Omega}=\Omega
[\sin ^2\theta +(\cos \theta -\frac i2\frac \kappa \Omega )^2]^{1/2}.
\label{cdphase}
\end{equation}
The imaginary part of $\tilde{\phi}_{\pm }^d$ describes the decay of the 
trajectory probability, a different decay rate is obtained for different 
basis states. This latter fact implies a distortion of the evolution 
respect to the dissipation-free process. 

More specifically for the desired geometric action, equation (\ref{cdphase})
reveals that the dissipation-induced distortion compared with $\phi ^g$ is
in an order of $\kappa /\gamma $, which for an adiabatic 
process with small $\gamma$, can be rather severe.
Nevertheless this distortion can be
avoided in the whole procedure since, by removing the dynamical phase, the
refocusing evolution with the opposite direction will remove the inhomogeneity 
of dissipation as well. The overall process gives merely origin to a global suppression 
of the trajectory probability with $\Gamma _d=e^{-\frac 12\kappa T}$. 

The cancellation of the unwanted non-unitary effects achieved by the double loop technique
is analog to a sort of noise decoupling by symmetrization via $\sigma_x$-pulses
\cite{bang}. Indeed the kinematical symmetry $\tilde{H}_0^{\bar{C}}(t)=\tilde{H}_0^C(T-t)$ 
of the control process, along with the exchange of the computational basis states 
over the second loop leads to a homogeneous, i.e., global action of the dissipation 
that can then be neglected. 

It should be noted that the effect of dissipation on the dynamical 
phase discussed above is not the only influence that dissipation has on the 
quantum dynamics of the system. Indeed, the non-unitary character of the 
evolution will affect the geometrical phase too. One is then lead to consider 
a complex Berry phase describing the dissipative effects embedded inside the 
geometric action,
\begin{equation}
\tilde{\phi}_{\pm }^g=\pm \pi [1-(\Omega /\bar{\Omega})(\cos \theta -\frac
i2\frac \kappa \Omega )]. 
\label{gphase}
\end{equation}
In view of the sensitivity of the geometric phase under the reversal of the 
orientation of the second control loop the dissipative effects
in Eq. (\ref{gphase}) cannot be removed via the refocusing process.
On the other hand, in case of low dissipation, equation (\ref{gphase}) reveals
that the resulting error is of order $\kappa /\Omega $ and thus can be
made negligible by making $\Omega$ large enough.

{\em Dissipation in holonomic quantum computation}.
We turn now to consider another well-established approach, namely, the so-called
holonomic quantum computation (HQC) \cite{zanardi}. In HQC information
is encoded in a degenerate eigenspace of the governing Hamiltonian. Pure
geometric operations, including the Abelian phase factor and
non-Abelian transformation, are naturally achieved by adiabatic
evolutions. Dynamical phase induced herein is clearly global
due to the inherent structural symmetry (degeneracy) of the
system levels. Dissipation occurring in this system can be
divided into two categories: (i) dissipation preserving the
degeneracy structure and (ii) inhomogeneous dissipation lifting the degeneracy.
It can be easily seen that in case (i) of homogeneous 
dissipation, only global decay can be induced. Hence the distortion to the 
evolution can be avoided in the no-jump trajectory \cite{note}. In what 
follows we shall focus on the general case of inhomogeneous dissipation, 
which will affect the cyclic evolution of the system in a more severe manner.

Consider as an illustration the standard optical system \cite{duan,recati}
in which a qubit is encoded in a multi-level $\Lambda $-type trapped ion or
a similar cavity atom. The ground (or metastable) levels $|g_i\rangle $ $
(i=1,\cdots ,n)$ are highly degenerated and each couples to the excited
state $|e\rangle $ in a tunable manner. Let the states $|g_1\rangle $ and 
$|g_2\rangle $ correspond to the computational basis $|0\rangle $ and $|1\rangle $,
respectively. In order to obtain the gate operation $e^{i\phi ^g|1\rangle \langle 1|}$, 
the system is described by the periodic Hamiltonian
\begin{equation}
H_C(\theta ,\varphi )=\Omega \sin \theta (\sigma _{2e}+\sigma _{e2})+\Omega
\cos \theta (\sigma _{3e}e^{i\varphi }+\sigma _{e3}e^{-i\varphi }),
\label{abeih}
\end{equation}
where $\sigma_{ei}:=|e\rangle\langle g_i|$ $(i=1,2,3)$,
and the parameters $\theta $, $\varphi $ evolve adiabatically. 
In the dissipation-free case the dark state of the system given by
$|D(\theta,\varphi )\rangle =\cos \theta |g_2\rangle -\sin \theta e^{i\varphi
}|g_3\rangle $, after a loop in the parameter space acquires a net 
Berry phase $\phi ^g$ which depends only
on the solid angle of the loop. Suppose now that the level $|g_3\rangle $
is metastable and suffers damping with $\Gamma ^{\dagger }\Gamma 
=\kappa_3\sigma _{33}$. One has that $\langle D(\theta ,\varphi )|
\kappa_3\sigma _{33}|D(\theta ,\varphi )\rangle =
\kappa_3\sin ^2\theta.$ The dissipative dynamics lifts the 
degeneracy between the dark state and the ground state $|g_1\rangle $. 
The evolution of the no-jump trajectory is described by the following 
non-unitary transformation 
\begin{equation}
|\psi (t)\rangle =\frac{u_C|\psi (0)\rangle }{\sqrt{\langle \psi
(0)|u_C^{\dagger }u_C|\psi (0)\rangle }}, \label{state}
\end{equation}
where the operator $u_C$ is given by
\begin{equation}
u_C\simeq e^{(\tilde{\phi}^d+i\phi ^g)|1\rangle \langle 1|}, 
\label{abeiu}
\end{equation}
with $\tilde{\phi}^d=-\frac 12\int_{0}^{T}\!\kappa_3\sin ^2\theta dt$. 
In detail, for the typical process in which $\varphi =\gamma t$ 
rotates uniformly, one has explicitly $\phi ^g=4\pi \sin ^2\theta $ 
and $\tilde{\phi}^d=-\pi (\kappa_3/\gamma )\sin ^2\theta $. 
Dissipation induced distortion in this process scales 
as $\kappa_3/\gamma $, which turns out to be unfavorable since for the adiabatic 
processing, the damping rate $\kappa_3$ may be comparable to the 
parameter frequency $\gamma $. Note also that in the above calculation, 
a low damping rate has been assumed: we have ignored the 
influence of dissipation on the geometric phase $\phi ^g$ and 
the deviation to the instantaneous dark state. Both of these effects would
cause an error to the gate operation scaling as $\kappa_3/\Omega.$ 
This latter ratio we assume can be made negligible.

Let us investigate further the influence of dissipation on the other single-qubit 
holonomic gate studied in \cite{duan}. We consider the adiabatic evolution 
generated by the Hamiltonian 
\begin{eqnarray}
H(t) &=&\Omega \sin \theta [\cos \varphi (\sigma _{1e}+\sigma _{e1})+\sin
\varphi (\sigma _{2e}+\sigma _{e2})] 
\nonumber \\
&\ &+\Omega \cos \theta (\sigma _{3e}+\sigma _{e3}).
\label{ham2}
\end{eqnarray}
The system admits two dark states
$|D_1\rangle =\cos \theta (\cos \varphi |g_1\rangle +
\sin \varphi |g_2\rangle )-\sin \theta |g_3\rangle $ 
and $|D_2\rangle =\cos \varphi |g_2\rangle -\sin \varphi |g_1\rangle $. 
A suitable cyclic evolution in the dissipation-free process
will generate the non-Abelian holonomic gate operation
$e^{i\phi ^g\sigma _y}$; the parameter $\phi ^g$ describes 
again the solid angle swept by the executed loop. We assume 
the same dissipation of the metastable level $|g_3\rangle $. 
It should be noted that the dissipation generated 
dynamics for the two dark states does not commute with 
the non-Abelian connection. In view that 
$\langle D_i|\kappa_3\sigma _{33}|D_j\rangle =
\kappa_3\sin ^2\theta $ for $i=j=1$ 
and null for other choices of $i$ and $j$, a perturbation
calculation gives directly that
\begin{equation}
u_C\simeq e^{\tilde{\phi}^d|0\rangle \langle 0|+
i\phi ^g\sigma _y}, 
\label{noabei}
\end{equation}
where $\tilde{\phi}^d=-\frac 12\int_{0}^{T}\!\kappa_3\sin ^2\theta dt$. 
For the specified process with $\varphi =\gamma t$, 
one has $\phi ^g=2\pi \cos \theta $ and
$\tilde{\phi}^d=-\pi (\kappa_3/\gamma )\sin ^2\theta $.
Similar to the former case, the distortion induced by inhomogeneous 
dissipation to this gate is also of order $\kappa_3/\gamma $.

{\em Generalized refocusing scheme for holonomic quantum computation}. 
We have shown in the above the influence of dissipation on
two different schemes for geometric quantum computation. 
At this stage, an interesting question arises: Is the refocusing 
strategy described in the former scheme extendable to the case of 
non-Abelian holonomic operations?
In what follows we will explicitly address that question for
the kind of optical system proposed in \cite{duan}. The 
answer will turn out to be ``yes".

It is worthwhile to stress the role of symmetry in the
dynamical evolutions of above systems.
The kinematical symmetry of refocusing scheme,
i.e., $H_{c}(t)=H_{\bar{c}}(T-t),$ helps in
quenching the dissipation-induced distortion. This is due to the fact
that refocusing removes the dynamical contributions from the evolution 
and these latter are the ones which are more severely affected by dissipation.
On the other hand, the symmetry associated with the level degeneracy
in the holonomic scheme is lifted by the dissipative interaction.
This in turn gives rise to an inhomogeneous decay of basis states
and to a corresponding distortion of the holonomic gates.
The scheme we are going to discuss in the following basically combines
the kinematical symmetry of the double-loop refocusing scheme with
a technique for restoring the dark-state symmetry and therefore homogeneity 
of the decay process. 

In brief the novel refocusing scheme for holonomic 
quantum computation that we are going to analyze, can be described as a 
double-loop configuration in which the additional loop refocuses the first 
one in an opposite direction. As one may see, the opposite loop execution 
would induce the inverse operation hence cancel the desired transformation, 
but this can be side stepped through interchanging the 
roles of the two computational bases in the twice loop evolutions.

Let us start by considering in a detailed fashion the single-qubit gate 
$e^{i\phi ^g|1\rangle \langle 1|}$; in the new scheme this gate can be achieved 
simply by performing the loop evolution twice, the first one as in 
(\ref{abeih}) followed by a second one described by 
\begin{equation}
H_{\bar{C}}(\theta ,\varphi )=\Omega \sin \theta (\sigma _{1e}+\sigma
_{e1})+\Omega \cos \theta (\sigma _{3e}e^{i\varphi }+\sigma
_{e3}e^{-i\varphi }),
\label{abeihh}
\end{equation}
where the parameters $\theta $ and $\varphi $ refocus the first loop 
but in the opposite direction. Note in the evolution of Eq. (\ref{abeihh}) 
the level $|g_1\rangle $ was employed to replace $|g_2\rangle $ 
in the first loop of Eq. (\ref{abeih}). 
We have interchanged the roles of the two basis states $|g_1\rangle $ and
$|g_2\rangle $ in order to make the overall dynamical evolution symmetric with respect
the exchange of states of the computational basis. This sort of symmetrization
plays the role of the two hard $\pi $ pulses used in the former NMR scheme to 
flip the basis states.

The dissipative dynamics 
on the second loop (\ref{abeihh}) will generate the transformation 
with
\begin{equation}
u_{\bar{C}}\simeq e^{(\tilde{\phi}^d-i\phi ^g)|0\rangle \langle 0|}.
\label{abeiuf}
\end{equation}
Thus the total two-loop process will be described by the following 
overall transformation
\begin{eqnarray}
u_ {dl}&\simeq &e^{(\tilde{\phi}^d-i\phi ^g)|0\rangle 
\langle 0|}\times e^{(\tilde{\phi}^d+i\phi ^g)|1\rangle \langle 1|}
\nonumber\\
&=&e^{\tilde{\phi}^d}e^{-i\phi ^g}e^{i2\phi ^g|1\rangle \langle 1|},
\label{abenew}
\end{eqnarray}
where $e^{-i\phi ^g}$ denotes an irrelevant global phase.
From this equation it can be clearly seen, in view of the symmetric role 
played by $|0\rangle$ and $|1\rangle $, that dissipation gives rise just 
to a global factor $e^{\tilde{\phi}^d}$ affecting just the no-jump 
trajectory probability; the former distortion in
equation (\ref{abeiu}) has been suppressed.

We now discuss the extension of the refocusing strategy
to the second single-qubit gate considered in \cite{duan}, i.e., 
$e^{i\phi ^g\sigma _y}$; this extension turns out to be more 
difficult than the previous one. Indeed, one may naively 
expect that the refocusing loop is described by 
\begin{eqnarray}
H_{\bar{C}}(\theta ,\varphi ) &=&\Omega \sin \theta 
[\cos \varphi (\sigma _{2e}+\sigma
_{e2})+\sin \varphi (\sigma _{1e}+\sigma _{e1})] 
\nonumber \\
&\ &+\Omega \cos \theta (\sigma _{3e}+\sigma _{e3}) 
\label{nonabeh}
\end{eqnarray}
with the parameter $\theta $ and $\varphi $ evolving in the opposite
direction of the first loop (\ref{ham2}). However, it is immediate to 
verify that the evolution associated to the loop (\ref{nonabeh}), 
namely the transformation $u_{\bar{C}}=e^{\tilde{\phi}^d|1\rangle 
\langle 1|+i\phi ^g\sigma _y}$,
does not commute with that of Eq. (\ref{noabei}). 
In other terms the non-commutativity of the 
two loop transformations destroys the desired symmetry aimed by
the refocusing process. As a result such a double-loop evolution 
does not offer any clear advantage as far as 
resilience against dissipation is concerned.
One can overcome this obstacle by performing the two opposite controlled 
evolutions in sense simultaneously. As the superposed process shall lead 
overlapped interactions to the level $|g_3\rangle $, an extra level will 
be necessary to pin it down. Explicitly, the whole process of the 
dual-superposed loop evolution can be described by the following periodic 
Hamiltonian
\begin{eqnarray}
H_{dl} &=&\Omega \sin \theta [(\cos \varphi -\sin \varphi )\sigma
_{1e}+(\sin \varphi +\cos \varphi )\sigma _{2e}]
\nonumber \\
&\ &+\Omega \cos \theta (\sigma _{3e}+\sigma _{4e})+h.c.. 
\label{nonnew}
\end{eqnarray}
Here the level $|g_3\rangle $ and the extra level $|g_4\rangle $ are 
supposed to suffer the homogeneous damping $\Gamma ^{\dagger }\Gamma 
=\kappa_3(\sigma _{33}+\sigma _{44})$. As the parameters $\theta $ and
$\varphi $ evolve adiabatically, the two dark states of the newly 
established Hamiltonian (\ref{nonnew}),
\begin{eqnarray}
|D_1^{dl}\rangle &=&\cos \theta (\cos \varphi |g_1\rangle +
\sin \varphi |g_2\rangle )-\sin \theta |g_3\rangle , 
\nonumber \\
|D_2^{dl}\rangle &=&\cos \theta (\cos \varphi |g_2\rangle -
\sin \varphi |g_1\rangle )-\sin \theta |g_4\rangle ,
\label{darknew}
\end{eqnarray}
shall generate a non-Abelian holonomy and it can implement the identical 
gate operation $e^{i\phi ^g\sigma _y}$ with $\phi ^g$ denoting the 
loop-related solid angle. The advantage of the present system is obvious:
the yielded new dark states possess completely symmetric structure 
which is thus capable to remove the inhomogeneous effect of dissipation 
on the computational space. Simply the dissipation can merely lead to a 
global decay and for the no-jump trajectory the transformation is described by
\begin{equation}
u_{dl}\simeq e^{\tilde{\phi}^d}e^{i\phi ^g\sigma _y}.
\label{nontrand}
\end{equation}
It should be pointed out that the loop superposition strategy is also 
realizable for the previously discussed single-qubit holonomic gate, 
which thus provides an alternative way to implement the gate 
$e^{i\phi ^g|1\rangle \langle1|}$. We stress here that the homogeneity 
achieved in the loop-superposed processes is restricted 
on the two-dimensional subspace spanned by the computational basis. 

In this paper we analyzed the robustness of geometric quantum gates, 
both Abelian and non-Abelian, against quantum dissipation. 
We have adopted the quantum jump approach to describe dissipative 
dynamics and we focused on the no-jump trajectory. It has been shown that 
geometric gates realized by means of a refocused double-loop scheme possess 
a certain resilience against dissipation-induced distortion.
This robustness is due to the underlying symmetry of the scheme with respect to 
the exchange of computational basis states. The crucial point is that the 
non-unitarity affects mostly the dynamical phases; these latter can be removed 
by a suitable control process which leaves the geometrical part of the 
evolution untouched. The power of this refocusing strategy has been demonstrated
by extending it to non-Abelian holonomic gates. For the non-Abelian case, we have 
shown how the same sort of symmetrization can be achieved via a loop superposition 
strategy. We conclude by noticing the central role of symmetry in a variety of 
schemes of quantum information processing (see e.g., Ref. \cite{symme} and references 
therein). The present work reveals one more intriguing way in which symmetry might 
help in our struggle against decoherence.

This work is financially supported by the European Union project TOPQIP (Contract 
No. IST-2001-39215).


\begin{thebibliography}{99}
\bibitem{note} In case of homogeneous dissipation, the trajectories 
with jumps occurring have also been discussed in Ref. \cite{guridi}.
\end{thebibliography}
\end{document}